\documentclass[aps,prb,twocolumn,floatfix,eqsecnum,showpacs,superscriptaddress]{revtex4}
\usepackage{graphicx}
\usepackage{amsmath}
\usepackage{amssymb}
\usepackage{txfonts}
\usepackage{bm}

\begin{document}

\title{Thermopower and the Mott formula for a Majorana edge state}
\author{Chang-Yu Hou}
\affiliation{Department of Physics and Astronomy, University of California
at Riverside, Riverside,
 CA 92521}
\affiliation{Department of Physics, California Institute of Technology,
Pasadena, CA 91125}
\author{Kirill~Shtengel}
\affiliation{Department of Physics and Astronomy, University of California
at Riverside, Riverside,
 CA 92521}
 \affiliation{Institute for Quantum Information, California Institute of Technology,
Pasadena, CA 91125}
\author{Gil~Refael}
\affiliation{Department of Physics, California Institute of Technology,
Pasadena, CA 91125}
\date{\today}

\begin{abstract}
We study the thermoelectric effect between a conducting lead and a Majorana
edge state. In the tunneling limit, we first use the Landauer-B\" uttiker
formalism to derive the Mott formula relating the thermopower and the
differential conductance between a conducting lead and a superconductor.
When the tunneling takes place between a conducting lead and a Majorana
edge state, we show that a non-vanishing thermopower can exist. Combining
measurements of the differential conductance and the voltage induced by the temperature difference between the conducting lead and the
edge state, the Mott formula provides a unique way to infer the temperature
of the Majorana edge state.
\end{abstract}

\maketitle

\section{Introduction}
\label{sec:intro}

Electron thermometry is a crucial component in most condensed matter experiments. It is especially necessary for exploring  the unconventional thermoelectric response of low-dimensional systems~\cite{Kane96,Kane97}. One
technique for probing the electron temperature utilizes the Seebeck effect by measuring the thermally induced voltage difference between a sample and a weakly coupled lead in the absence of a current. Then, using the Mott formula~\cite{Mott,Jonson80,Sivan86},
\begin{equation}
\label{eq:Mott-formula}
\mathcal{S} =-\frac{\Delta V}{\Delta T}= \frac{\pi^2}{3}
\left( \frac{k_B^2 T}{e}\right) \left( \frac{d \ln G (E)}{d E} \right)_{E= \mu},
\end{equation}
the temperature of the sample can be inferred from the differential
conductance, $G(E)$, at energy $E$. Here, $\mathcal{S}$ is the thermopower
(Seebeck coefficient) defined as the ratio of the voltage difference, $\Delta V$, and the temperature difference, $\Delta T$, between the sample and the lead, while $T$ and $\mu$ can be taken as the average temperature and chemical potential, respectively. Experimentally, such a technique was first demonstrated in quantum point-contact devices~\cite{Molenkamp90,Appleyard98} and later used to measure the temperature variation of quantum Hall edge
states~\cite{Granger09}.

In this paper, we consider thermal and electric transport between a
conducting lead and a Majorana edge state that appears at the boundary of a
two-dimensional chiral $p$-wave superconductor~\cite{Read00}. It is not
clear, a priori, whether a non-vanishing thermopower can be established, since the Majorana edge mode is charge-neutral due to its underlying particle--hole symmetry. Here, we use the Landauer--B\"{u}ttiker formalism~\cite{Buttiker86,Datta,Blonder82,Claughton96,Anantram96} and show
that the Mott formula for the thermopower between a superconducting sample
and a conducting lead is satisfied in general once both normal and Andreev
scattering processes are taken into account.~\cite{note1} We will then focus on several simplified models in order to address the utility of the Mott formula for probing the temperature of a Majorana
edge state. This technique could be naturally used in p+ip superconductors to probe the non-Abelian nature of Majorana zero modes through their unique magneto-thermoelectric signatures~\cite{Hou12}.

The paper is organized as follows.
In Sec.~\ref{sec:QD}, we consider a setup in which a quantum dot with discrete quantum levels couples weakly to a Majorana edge state. As a proof of principle, we show that the thermoelectric response in such setups can be non-vanishing. In the absence of a current, a finite voltage is established in the presence of a temperature difference between the quantum dot and the edge state.
In Sec.~\ref{sec:scattering}, the linear thermoelectric response coefficients
between a metallic and a superconducting lead are expressed in terms
of scattering probabilities within the framework of the Landauer-B\"uttiker
formalism. The Mott formula is then derived from the response coefficients.
In Sec.~\ref{sec:1-channel}, we explicitly derive the scattering matrix for a single-channel lead coupled to a Majorana edge state. Two scenarios are considered: (a) a single point-contact and (b) a
double point-contact. We demonstrate that the single point-contact
setup has vanishing thermopower, while the double point-contact setup generically has non-vanishing thermopower. We also discuss the possible thermopower strength for case (b).
We conclude our paper in Sec.~\ref{sec:conclusion}. To supplement discussions in the main text, we also include two Appendices that provide a proof of the Onsager relation and a list of the scattering matrix elements for a double-point-contact setup.

\section{Coupling between a Quantum dot and the Majorana edge state}
\label{sec:QD}

\begin{figure}
\includegraphics[angle=0,scale=0.9]{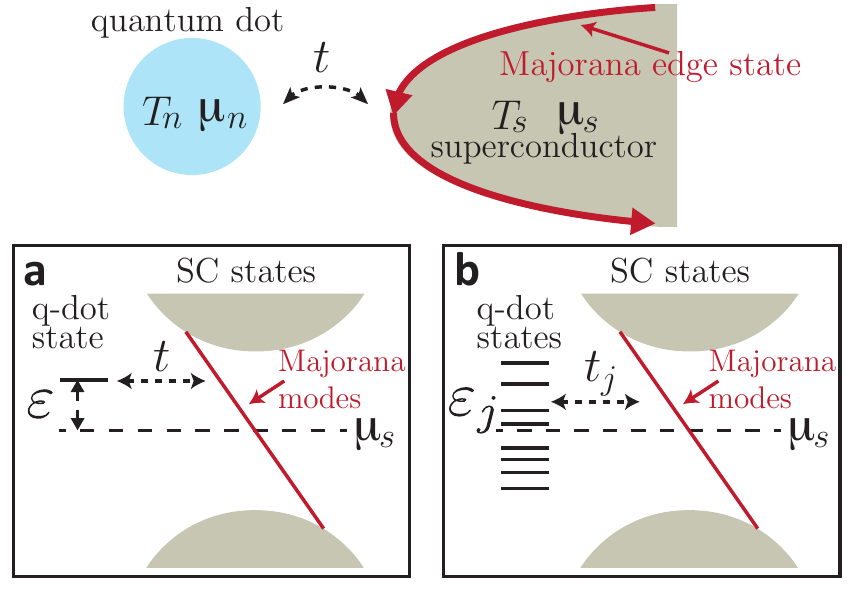}
\caption{The upper panel schematically depicts a quantum dot weakly
 coupling to a Majorana edge state. The temperature and chemical potential of
 the quantum dot are $T_n$ and $\mu_n$ respectively, and those of the Majorana
 edge state are $T_s$ and $\mu_s$. (a) A quantum dot with a single
 energy level $\varepsilon$ is coupled to a Majorana edge state
 with tunneling strength $t$. (b) Multiple
 energy levels $\varepsilon_j$ of a quantum dot are coupled to
 a Majorana edge state with tunneling strengths $t_j$ for each state.
}
\label{fig:q-dot}
\end{figure}

To gain some intuition as to how a non-vanishing thermopower can arise between a conducting lead and the Majorana edge state, we begin by considering a simplified model in which a conducting lead is replaced by a non-superconducting
quantum dot, as schematically shown in Fig.~\ref{fig:q-dot}.

\subsection{Single-state quantum dot}

Let us first consider the case in which a quantum dot consisting of a single
quantum state weakly couples to the chiral Majorana edge state as shown in
Fig.~\ref{fig:q-dot}a. The quantum dot has temperature $T_\text{n}$ and chemical potential $\mu_\text{n}$, while the chiral Majorana edge state
has temperature $T_\text{s}$ and chemical potential $\mu_\text{s}=0$.
All energies are measured with respect to the chemical potential of the
superconductor.

In the continuum limit, the effective Hamiltonian reads
\begin{equation}
\label{eq:H-single-state}
H = \varepsilon c^{\dag} c + i \frac{v_\text{m}}{2} \int \eta (x) \partial_x \eta(x) dx + i t (c + c^{\dag}) \eta(0)
\end{equation}
where $c$ is the annihilation operator of the fermionic state with energy
$\varepsilon$ in the quantum dot~\cite{Reimann02}, $\eta(x)$ represents the chiral Majorana
fermion mode~\cite{Read00, Law09}, $v_\text{m}$ is the velocity of the edge state, and the
coordinate $x$ runs along the boundary of the superconductor.~\cite{note2} The last term describes the fermionic coupling, with strength $t$, between the quantum dot and the Majorana edge state at $x=0$. The Majorana edge state appears at the boundary of a two-dimensional chiral $p$-wave superconductor, which reflects the topological property of the superconductor~\cite{Read00}. As a result of the spontaneously broken time-reversal symmetry, it flows with a definite chirality along the edge, akin to the quantum Hall edge state. However, since the Majorana edge state is Bogoliubov-de Gennes quasiparticles, it consists of half of the quantum Hall edge state degrees of freedom. Due to the particle-hole symmetry, the Majorana edge state is formally expressed as a chiral real (Majorana) field, $\eta(x)$.

The Majorana edge state can be represented in momentum space by
\begin{equation}
\eta(x) = \int_{-\infty}^{\infty} \frac{dk}{2\pi} e^{-i k x} \eta(k),
\end{equation}
where $\eta^{\dag}(k)=\eta(-k)$ due to the real nature of the field,
$\eta^{\dag} (x)=\eta (x)$. Under the transformation, we have
\begin{multline}
H = \varepsilon c^{\dag} c +   \int_{k>0} \frac{dk}{2\pi}  \varepsilon_k \eta^{\dag} (k) \eta(k)
\\
+ i t (c + c^{\dag}) \int_{k>0} \frac{dk}{2\pi} (\eta(k)+\eta^\dag(k)),
\label{eq:H-single-state-1}
\end{multline}
where the spectrum of the edge state is given by $\varepsilon_k= v_\text{m} k$
and we have used the relation $\eta^\dag(k)= \eta(-k)$ for the last term.Note that the symbol $\varepsilon$ is used to denote the energy spectrum of either the quantum dot or the edge modes. Now, we can treat the first line of Eq.~\eqref{eq:H-single-state-1} as the unperturbed Hamiltonian $H_0$, and the second line as the perturbed term $V$.

The transition rate between the quantum dot and the edge state is given by the Fermi-golden rule,
\begin{equation}
\label{eq:Fermi-golden-rule}
 T(i\to f)= \frac{2\pi}{\hbar} |\langle f | V | i \rangle|^2 \delta (E_i-E_f),
\end{equation}
where $i$ and $f$ indicate the initial and the final states of a tunneling process with the corresponding total energies $E_{i}$ and $E_f$ of the system, respectively, and
the delta function enforces the energy conservation. To compute the matrix
elements, let us denote the states of the system as $|n_c n_{\eta(k)}
\rangle$, where $n_c,n_{\eta(k)}=0,1$ represent occupation numbers of the
quantum dot state and chiral edge states with energy $\varepsilon_k$
respectively. With the assumption $\varepsilon>0$, energy conservation implies that fermions can tunnel between states with energy
$\varepsilon_k=\varepsilon$ for single particle processes. Only two matrix
elements $\langle 0 1| V | 10\rangle = \langle 10| V | 01 \rangle = i t$ are
non-vanishing and give tunneling rates
\begin{align}
 T(10 \to 01) =T(01 \to 10)= \frac{2\pi}{\hbar} t^2 \delta (\varepsilon- \varepsilon_k).
\end{align}

With these tunneling rate, the current tunneling from the quantum dot to the
edge state can be written as
\begin{equation}
\label{eq:current-single-state}
 I  =  - \frac{e}{\hbar} \frac{t^2}{v^{\ }_\text{m}} \left[f_n(\varepsilon) - f_s(\varepsilon) \right],
\end{equation}
where $f_{n}(E)$ and $f_{s}(E)$ are the Fermi-Dirac distributions of the quantum dot and the edge state, respectively. To first order in $\mu_n-\mu_s$ and $T_n-T_s$, we obtain a linearized current
response
\begin{equation}
\label{eq:current-single-state-1}
I =  \frac{ e}{\hbar} \frac{t^2}{v^{\ }_\text{m}}
\left(\frac{\partial f_\text{s}(E)}{\partial E}\right)_{E=\varepsilon}
\left[ (\mu_n-\mu_s)  + \frac{\varepsilon}{T_s} (T_n-T_s) \right],
\end{equation}
that generically has a thermoelectric response such that the temperature
difference will lead to a potential difference with a vanishing current and
hence a non-vanishing thermopower.

\subsection{Multilevel quantum dot}

For a quantum dot whose energy spectrum contains multiple levels as shown in
Fig.~\ref{fig:q-dot}b, the Hamiltonian can be modeled as $\sum_{j}
\varepsilon_{j} c_j^{\dag} c_j $, where $c_j$ is the annihilation operator
for the state with energy $\varepsilon_{i}$. Now, the Hamiltonian
representing the coupling between the quantum dot states and the Majorana
edge state can be generically written as
\begin{equation}
\label{eq:V-few-state}
V=+ i \sum_{j} t(\varepsilon_j) (c_j + c_j^{\dag}) \int_{k>0} (\eta(k)+\eta^\dag(k)),
\end{equation}
where $t(\varepsilon_j)$ is the coupling strength for each energy $\varepsilon_j$.

In the weak tunneling limit, i.e., where higher order scattering processes are
neglected, the linearized current flowing from the quantum dot to the edge
state becomes
\begin{equation}
I = \frac{e}{\hbar} \sum_{j} \frac{ t(\varepsilon_j) ^2}{v^{\ }_\text{m}}
\left(\frac{\partial f_{\text{s}}(E)}{\partial E}\right)_{E=\varepsilon_j}
\left[  (\mu_n-\mu_s)   + \frac{\varepsilon_{j}}{T_s} (T_n-T_s) \right].
\end{equation}
Now, a finite thermoelectric response appears when the summation of the second term is finite. In the continuum limit, we can define the density of states of the quantum dot as $\rho(E)$ and take $\varepsilon_j\to E$ to the
continuum energy. Because $E \left({\partial f(E)}/{\partial E} \right)$ is
an odd function of energy $E$, we need $\rho(E) t(E)^2 $ to be non-even in order for the thermopower not to vanish. Hence we conclude that the asymmetry of
$\rho(E) t(E)^2$ as a function of energy is crucial for obtaining a finite
thermopower.

Since only lowest-order tunneling processes are considered throughout this
section, we have neglected higher-order scattering processes present at NS
junctions, specifically Andreev scattering. To include contributions of
Andreev scattering processes, we will employ the Landauer--B\" uttiker
formalism for scattering between the conducting lead and the superconductor
in the next section.

\section{Landauer--B\" uttiker Scattering Formalism}
\label{sec:scattering}

As we have shown in the previous section, the thermopower at the boundary
between a quantum dot and a Majorana edge state is generically non-vanishing.
Here, by using the Landauer--B\" uttiker formula, we derive a Mott formula
for the thermopower between a conducting lead and a superconducting region,
akin to the normal Mott formula in Eq.~\eqref{eq:Mott-formula}.

\begin{figure}
\includegraphics[angle=0,scale=0.9]{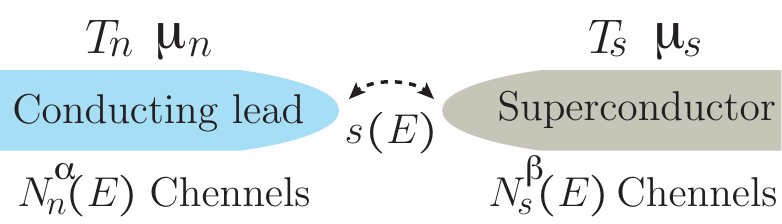}
\caption{Schematic picture of a tunneling junction between a conducting lead and
a superconductor. The conducting lead has $N_{n}^{\alpha}(E)$ channels at
energy $E$ while the superconductor has $N_{s}^{\beta}(E)$, where
$\alpha (\beta)=p,h$ represents the quasiparticle or quasihole channels.
The temperature and chemical potential of the conducting lead are $T_n$ and
$\mu_n$ respectively, and those of the superconductor are $T_s$ and $\mu_s$.
$s(E)$ represents the scattering matrix across the tunneling junction.
}
\label{fig:general-setup}
\end{figure}

Let us consider a generic setup shown in Fig.~\ref{fig:general-setup}, in which
a conducting lead weakly couples to a superconductor lead. The temperature
and chemical potential of the conducting lead are $T_n$ and $\mu_n$,
respectively, and those of the superconductor are $T_s$ and $\mu_s$. Again,
we set $\mu_s=0$ in what follows. To adapt the Landauer--B\" uttiker formula,
we describe both the conducting lead and the superconductor by
one-dimensional channels. Since a quasiparticle can scatter into either a
quasiparticle or a quasihole, we need to specify numbers of quasiparticle
($p$) and quasihole ($h$) channels. At a given energy $E$, we denote
$N_{n\alpha}(E)$ and $N_{s \alpha}(E)$ as the number of $\alpha=p/h$ channels
of the conducting lead and the superconductor, respectively. We note that
quasiparticles are electrons in the conducting lead while in the
superconductor they are Bogolyubov--de Gennes quasiparticles.

In the weak tunneling and dc limits, all transport properties between the
conducting lead and the superconductor are governed by a unitary scattering
matrix. The scattering matrix elements, which give scattering amplitudes from
channel $(j,\beta,b)$ to $(\ell,\alpha,a)$, are denoted by $s^{i \alpha a}_{j
\beta b} (E)$. Here, $\ell ,j=n/s$ are lead indices, $\alpha,\beta= p/h$ are
particle/hole indices, and $a= 1,\dots, N_{\ell \alpha}(E)$ ($b= 1,\dots,
N_{j\beta}(E)$) are channel indices. We adopt the convention that lower
indices represent an incoming state while the upper indices represent an
outgoing state. The particle--hole symmetry of superconductors gives
\begin{equation}
\label{eq:s-matrix-relation-PH-sym}
s^{\ell \alpha a}_{j\beta b} (E) = \alpha \beta [s^{\ell \bar{\alpha} a }_{j \bar{\beta} b} (-E)]^*,
\end{equation}
where particle indices $\alpha,\beta=p/h$ are defined as $+/-$ in the
equation and $\bar{\alpha},\bar{\beta}$ are defined as $\bar{p},\bar{h}\equiv
h,p$.

Now, the probability of an incoming current at the channel $(j,\beta,b)$
scattering into the outgoing current at the channel $(\ell,\alpha,a)$ is
given by $|s^{\ell \alpha a}_{j \beta b} (E)|^2$. As we are interested in the
total tunneling current between the conducting lead and the superconductor,
it is convenient to trace out channel indices and introduce the scattering
probability
\begin{equation}
\label{eq:def-prob-S}
P^{\ell\alpha}_{j\beta}(E) = \sum_{a,b} |s_{j \beta b}^{\ell \alpha a}(E)|^2
\end{equation}
for the current in the $(j,\beta)$ state to scatter into the $(\ell,\alpha)$
state. Here, $\ell=j$ corresponds to reflection while $\ell \neq j$
correspond to transmission. The processes with $\alpha \neq \beta$ are
Andreev scattering processes. Due to the unitarity of the scattering matrix,
scattering probabilities satisfy
\begin{equation}
\label{eq:unitarity-P-N}
 \sum_{j,\beta} P_{j \beta}^{\ell \alpha}(E)= N_{i \alpha} (E),
 \quad  \sum_{\ell,\alpha} P_{j \beta}^{\ell \alpha}(E)= N_{j \beta} (E).
\end{equation}
The particle--hole symmetry, Eq.~\eqref{eq:s-matrix-relation-PH-sym}, further
implies that
\begin{equation}
\label{eq:P-H-relation-P}
P_{j  \bar{\beta}}^{\ell \bar{\alpha}} (-E) =  P_{j \beta}^{\ell \alpha} (E).
\end{equation}
Combining Eqs.~\eqref{eq:unitarity-P-N} and \eqref{eq:P-H-relation-P}, we
find that the number of channels satisfies $N_{\ell h }(E)=N_{\ell p}(-E)$.

Now, electric and heat currents can be expressed in terms of scattering
probabilities~\cite{Blonder82,Claughton96,Anantram96}. The electric current
reads
\begin{multline}
\label{eq:electric-current}
I=  \frac{e}{h} \int_0^{\infty} d E \Big\{ - f^{p}_n(E)
\left( N_{np} - P_{np}^{np} + P_{np}^{nh} \right)
\\
+  f^{h}_n(E) \left( N_{nh} - P_{nh}^{nh} + P_{nh}^{np} \right)
\\
+ f_s^{p}(E) \left(P_{sp}^{np} - P_{sp}^{nh} \right)
+  f_s^{h}(E) \left(P_{sh}^{nh} -P_{sh}^{np} \right) \Big\},
\end{multline}
while the heat current reads
\begin{multline}
\label{eq:heat-flux}
Q=  \frac{1}{h} \int_0^{\infty} d E\, E \Big\{+ f^{p}_n(E)
\left( N_{n p} - P_{np}^{np} - P_{np}^{nh} \right)
\\
+ f^{h}_n(E) \left( N_{nh} - P_{nh}^{nh}- P_{nh}^{np}\right)
\\
- f_s^{p}(E) \left(P_{sp}^{np} + P_{sp}^{nh} \right) - f_s^{h}(E)
\left(P_{sh}^{np} + P_{sh}^{nh} \right)\Big\},
\end{multline}
where energy dependences of all $N_{\ell \alpha}$ and $P_{j\beta}^{\ell
\alpha}$ are implied. The Fermi-Dirac distributions of particles and holes
are given by
\begin{equation}
f^{\alpha}_{\ell} (E) = ( e^{(E \pm \mu_\ell)/k_\text{B} T_\ell} +1)^{-1},
\end{equation}
where $\ell=n/s$ and $\alpha=p/h$. The ``$+$'' sign in the
exponent corresponds to particles ($\alpha = p$), the ``$-$'' sign -- to
holes ($\alpha=h$). Since the particle--hole picture effectively maps
quasiparticles with negative energy to quasiholes with positive energy, the
integration in Eqs.~\eqref{eq:electric-current} and~\eqref{eq:heat-flux} is
performed only  over the positive energy region to avoid double counting. Intuitively, the expression of the electric current in
Eq.~\eqref{eq:electric-current} sums over flows in all channels of the
conducting lead multiplied by the sign of their charge carriers
(negative(positive) charge for the electron(hole)) while the heat flow expression
in Eq.~\eqref{eq:heat-flux} sums over energy flows in all channels.

With the aid of the unitarity properties~\eqref{eq:unitarity-P-N} and the
particle--hole symmetry~\eqref{eq:P-H-relation-P}, the electric current
becomes
\begin{equation}
\label{eq:electric-current-1}
I= \frac{-e}{h} \int_{-\infty}^{\infty} dE  \left( f_{n}^{p} - f_{s}^{p} \right)
\left( N_{np} -P_{np}^{np} + P_{np}^{nh} \right),
\end{equation}
while the heat current can be rewritten as
\begin{equation}
\label{eq:thermal-current-1}
Q= \frac{1}{h} \int_{-\infty}^{\infty} dE E \left( f_{n}^{p} - f_{s}^{p} \right)
\left( N_{np} -P_{np}^{np} - P_{np}^{nh} \right).
\end{equation}
Here, we have used the relation $f_n^{h}(E)=1-f_n^{p}(-E)$ to extend the
range of integration to all energies. Importantly, both electric and heat
currents involve only reflection probabilities in the conducting
leads.~\cite{Blonder82}

To linear order, the expressions for the electric and heat currents can be
organized in terms of the chemical potential difference $\Delta \mu
=\mu_n-\mu_s$ and the temperature difference $\Delta T =T_n-T_s$ as
\begin{equation}
\left(
\begin{array}{c}
I/(-e)
\\
Q
\end{array}
\right)
=
\left(
\begin{array}{cc}
L_{11} & L_{12}/T_s
\\
L_{21} & L_{22}/T_s
\end{array}
\right)
\left(
\begin{array}{c}
\Delta \mu
\\
\Delta T
\end{array}
\right).
\end{equation}
The linear response coefficients $L_{ij}$ are given by
\begin{subequations}
\label{eq:def-Lij}
\begin{align}
L_{11} =& \frac{-1}{h} \int_{-\infty}^{\infty} d E  \frac{\partial f}{\partial E} \left( N_{np} - P_{np}^{np} + P_{np}^{nh} \right),
\\
L_{12} =& \frac{-1}{h} \int_{-\infty}^{\infty} d E E  \frac{\partial f}{\partial E}   \left( N_{np} - P_{np}^{np} + P_{np}^{nh} \right),
\\
L_{21} =& \frac{-1}{h} \int_{-\infty}^{\infty} d E E  \frac{\partial f}{\partial E}  \left( N_{np} - P_{np}^{np} - P_{np}^{nh} \right) ,
\\
L_{22} =& \frac{-1}{h} \int_{-\infty}^{\infty} d E E^2  \frac{\partial f}{\partial E}   \left( N_{np} - P_{np}^{np} - P_{np}^{nh} \right) .
\end{align}
\end{subequations}
As shown in Appendix~\ref{app:Onsager-relation}, the Onsager reciprocal relation is satisfied for these linear response coefficients. In the presence
of time-reversal symmetry, one can show that $L_{12}= L_{21}$ even though the
expressions given by Eqs.~\eqref{eq:def-Lij} do not directly reflect this. In the absence of time-reversal symmetry, linear-response coefficients between
the system and its time-reversed system are related due to the Onsager relation,
i.e., $L_{12}=\mathcal{T}[L_{21}]$, where $\mathcal{T}[\dots]$ represents the coefficient in the time-reversed system.

As we are interested in the relation between $L_{11}$ and $L_{12}$, it is
convenient to define a function
\begin{equation}
\label{eq:K}
K(E)= N_{np}(E) - P_{np}^{np}(E) + P_{np}^{nh}(E).
\end{equation}
When the tunneling probabilities are smooth functions of energy $E$, the
linear response coefficients $L_{ij}$ can be approximated by Sommerfeld
expansions.~\cite{Ashcroft} To the lowest non-vanishing order, the Sommerfeld
expansion of $L_{11}$ is given by
\begin{equation}
\label{eq:L11-expansion}
L_{11} \approx  \frac{1}{h} K(E)\Big|_{E=0} + O(k_{\text B} T_s)^2,
\end{equation}
while the Sommerfeld expansion of $L_{12}$ reads
\begin{equation}
\label{eq:L12-expansion}
L_{12} \approx  \frac{1}{h} \frac{\pi^2}{3} (k_{\text B} T_s)^{2} \frac{dK(E)}{dE}
\Big|_{E=0}+  O(k_{\text B} T_s)^4 .
\end{equation}
We observe that both $L_{11}$ and $L_{12}$ can be related to the differential
conductance by $G(E=eV)=-\frac{e^2}{h} K(E)$ with an applied voltage $V$ at
the normal lead and with a fixed chemical potential at the superconductor.

Now, the thermopower (Seebeck coefficient) can be readily evaluated and shown
to satisfy the Mott formula
\begin{equation}
\label{eq:Seebeck-coeff}
 \mathcal{S}=-\frac{\Delta V}{\Delta T }=\frac{1}{e T_s} \frac{L_{12}}{L_{11}}
 =  \frac{\pi^2}{3} \frac{k_{\text B}^2 T_s}{e} \frac{d \ln G(E)}{dE}\Big|_{E=\mu_s} .
\end{equation}
Here, $\Delta V$ is the voltage in response to the temperature
difference $\Delta T$ between the tunneling junction in the absence of total
current. The thermopower vanishes when the differential conductance is an
even function of energy $E$. The presence of the Mott relation provides a
unique way to infer the temperature difference by measuring the differential
conductance and the voltage difference between the conducting lead and the
superconductor.

\section{Single-channel continuum models}
\label{sec:1-channel}

In this section, we discuss two simplified models that describe a single-channel conducting lead tunneling into a chiral Majorana edge state. These
models provide examples in which the thermopower is (A) vanishing as
tunneling probabilities have no energy dependence due to an accidental
symmetry for a single-point contact setup, or (B) non-vanishing as tunneling
probabilities gain energy dependence in a double-points contact setup. The
goal here is not to present realistic models for such systems, but rather to
argue that the scenario (B) represents a generic case for a realistic system.

\begin{figure}
\includegraphics[angle=0,scale=0.9]{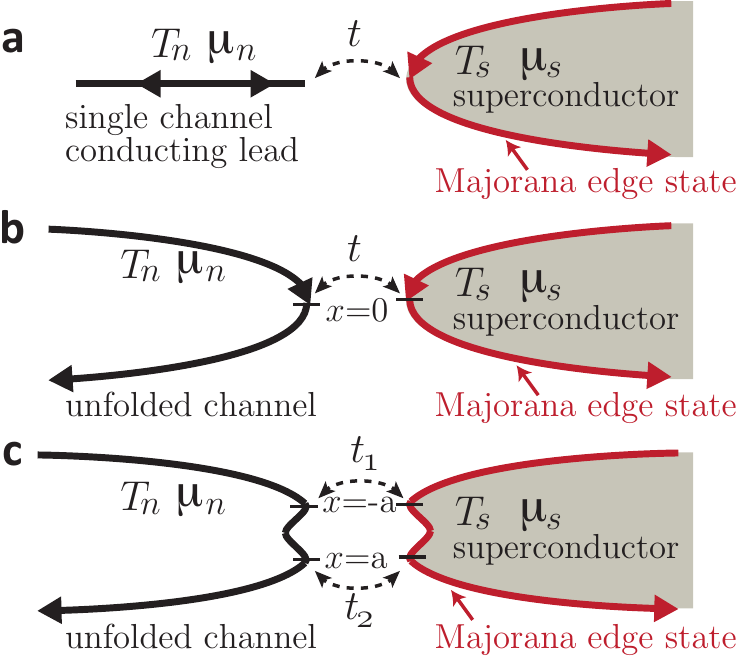}
\caption{(a) Schematic plot of a tip of a single-channel lead coupled to a Majorana
edge state (red curve at the boundary of the superconductor).
The tunneling strength is $t$. The temperature and chemical potential of
the conducting lead are $T_n$ and $\mu_n$, respectively, and those of the
Majorana edge state are $T_s$ and $\mu_s$. (b) A single-electron channel
can be unfolded into a chiral electron channel that couples to
a Majorana edge state at the point $x=0$. (c) The plot shows a toy model of
a double-point-contact setup between an unfolded electronic channel and
a Majorana edge state. Two tunneling points are at $x=\pm a$
along the chiral electron mode.
}
\label{fig:E-M-tunneling}
\end{figure}

\subsection{Single-point-contact geometry}

Let us consider a setup where a tip of a single-channel conducting lead
couples to a chiral Majorana edge state. In such a case, the non-chiral
channel can be unfolded to form a chiral electron mode as depicted in
Fig.~\ref{fig:E-M-tunneling}b. The coupling Hamiltonian is given
by~\cite{Law09,Li12}
\begin{equation}
\label{eq:chiral-electron-MF-tunneling}
\begin{split}
H =& H_N + H_{MF} + H_{t},
\\
H_N =& -i v_\text{f} \int_{-\infty}^{\infty} dx \psi^{\dag} (x) \partial_x \psi(x) ,
\\
H_{MF} = & -i \frac{v_\text{m}}{2} \int_{-\infty}^{\infty} dx \eta(x) \partial_x \eta(x) ,
\\
H_t = &  \frac{i}{\sqrt{2}} \int_{-\infty}^{\infty} dx \left( t \psi(x) + t^* \psi^{\dag} (x) \right) \eta(x) \delta(x) .
\end{split}
\end{equation}
Here, the $\psi(x)$ is the annihilation operators for the chiral electron
mode with velocity $v_\text{f}$, $\eta(x)$ is the chiral Majorana mode with
velocity $v_\text{m}$, and $t$ is the coupling strength. $\psi$ satisfies the
usual fermionic commutation relations, while the Majorana fermion satisfying
the anticommutation relation $\{\eta (x),  \eta (x')\} = \delta (x-x')$. The
equations of motion are readily written as
\begin{align}
\partial_t \psi =& -v_\text{f} \partial_x \psi+ t^* \eta \delta(x)/\sqrt{2},
\\
\partial_t \psi^{\dag} =& - v_\text{f} \partial_x \psi^\dag + t \eta \delta(x)/\sqrt{2},
\\
\partial_t \eta =& - v_\text{m} \partial_x \eta - \left(t \psi + t^* \psi^{\dag} \right) \delta(x)/\sqrt{2}.
\end{align}
By removing the time dependence with the Ansatz
\begin{equation}
\psi= e^{-i E t} \psi(x), \; \psi^{\dag}= e^{-i E t} \psi^{\dag}(x), \; \eta=e^{-i E t} \eta(x),
\end{equation}
we have
\begin{equation}
\begin{split}
 v_\text{f} \partial_x \psi(x) =& i E \psi(x) + t^* \eta (x) \delta(x)/\sqrt{2},
\\
 v_\text{f} \partial_x \psi^\dag(x) =& i E \psi^{\dag}(x)  + t \eta (x) \delta(x)/\sqrt{2},
\\
 v_\text{m} \partial_x (x) =& i E \eta(x)  - \left(t \psi(0)+ t^* \psi^{\dag} (x) \right) \delta(x)/\sqrt{2}.
\end{split}
\end{equation}

Now, a single-point-contact scattering problem at $x=0$ can be solved by the
transfer matrix method. First, the delta function $\delta (x)$ is
approximated as a bump function with width $\ell$ and height $1/\ell$. Then,
the transfer matrix of this bump geometry can be obtained by partitioning the
width $\ell$ by $N \to \infty$ steps. Finally, the $\delta(x)$ function is
recovered by taking the limit $\ell \to 0$. By following these steps, we
connect operators at $x=0^{+}$ to operators at $x=0^{-}$ by a transfer matrix
$M$ as
\begin{equation}
\left(
\begin{array}{c}
\eta(0^{+})
\\
\psi(0^{+})
\\
\psi^\dag(0^{+})
\end{array}
\right)
=
M
\left(
\begin{array}{c}
\eta(0^{-})
\\
\psi(0^{-})
\\
\psi^\dag(0^{-})
\end{array}
\right).
\end{equation}

As this transfer matrix $M$ gives the particle hopping amplitudes but not the current scattering amplitudes between the conducting lead and the Majorana
edge state, it is convenient to convert the transfer matrix to the scattering
matrix associating with the current scattering by including the effect of
different velocities. We obtain the scattering matrix
\begin{align}
\label{eq:S-matrix-single}
&S^t=
\left(
\begin{array}{ccc}
s_{\eta}^{\eta} & s_{p}^{\eta} & s_{h}^{\eta}
\\
s_{\eta}^{p} & s_{p}^{p} & s_{h}^{p}
\\
s_{\eta}^{h} & s_{p}^{h} & s_{h}^{h}
\end{array}
\right)
\\
&
=
\left(
\begin{array}{ccc}
\cos 2 \lvert \tilde{t}\rvert
&
-\frac{e^{i \phi}}{\sqrt{2} } \sin 2\lvert \tilde{t}\rvert
&
-\frac{e^{-i \phi}}{\sqrt{2} } \sin 2 \lvert \tilde{t}\rvert
\\
\frac{e^{-i \phi}}{\sqrt{2} } \sin 2 \lvert \tilde{t}\rvert
&
\cos^2 \lvert \tilde{t}\rvert
&
-e^{ -2 i \phi} \sin^2 \lvert \tilde{t}\rvert
\\
\frac{e^{i \phi}}{\sqrt{2} } \sin 2 \lvert \tilde{t}\rvert
&
-e^{2i \phi}\sin^2 \lvert \tilde{t}\rvert
&
\cos^2 \lvert \tilde{t}\rvert
\end{array}
\right),
\nonumber
\end{align}
where the effective coupling constant $\tilde{t}$ is defined as
$\tilde{t}={t}/{\left(2 \sqrt{v_\text{f} v_\text{m}}\right)}$ with $\phi$
being its phase: $t=|t|e^{i\phi}$. The $s^{\alpha}_{\beta}$ are matrix
elements corresponding to scattering amplitudes of a quasiparticle type
$\beta$ scattering to type $\alpha$. Here, $\eta$, $p$ and $h$ indicate
Majorana edge mode, electron and hole. One can directly verify that this
scattering matrix $S$ is unitary.

From Eq.~\eqref{eq:def-prob-S}, we conclude that the scattering probabilities
have no energy and phase dependence. Thereby, the kernel defined in
Eq.~\eqref{eq:K} and, consequently, the differential conductance will not
depend on the energy of the incoming flows. It then follows from the Mott
relation~\eqref{eq:Seebeck-coeff} that the thermopower vanishes for the
single-point-contact geometry. This is a consequence of uniform coupling
strengths between the states with different energies along with the constant
density of states of both electronic and Majorana modes -- an artifact of our
model.

\subsection{Double-point-contact geometry}

To emulate the finite extent of the tunneling region between a conducting leadand a Majorana edge state, we consider a double-point-contact setup. Once again, for simplicity, the non-chiral channel in the conducting lead is unfolded to form a chiral electronic mode. With two point contacts at $x= \pm a/2$, as shown in Fig. 3(c), the tunneling Hamiltonian $H_t$ becomes
\begin{multline}
H_{tt}=  \frac{i}{\sqrt{2}}
\left( t_1 \psi(-a/2) + t_1^* \psi^{\dag} (-a/2) \right) \eta(-a/2)
\\
+ \frac{i}{\sqrt{2}}  \left( t_2 \psi(a/2)
+ t_2^* \psi^{\dag} (a/2) \right) \eta(a/2),
\end{multline}
The scattering matrix $S_{tt}$ which connects currents on both sides of the
scattering region can be obtained by combining two scattering matrices in
Eq.~\eqref{eq:S-matrix-single} with the transfer matrix, and is given by
\begin{equation}
S^{tt}=
S (t_2)
\left(
\begin{array}{ccc}
e^{i k_m a} & 0 &0
\\
0 & e^{i k_f a} & 0
\\
0 & 0 & e^{i k_f a}
\end{array}
\right)
S (t_1).
\end{equation}
Here momenta are defined by $k_f=E /v_\text{f}$ and $k_m=E /v_\text{m}$ and
the tunneling amplitudes are defined by $t_1=|t_1|e^{i \phi_1}$ and $t_{2}=
|t_2| e^{i \phi_2}$.

As the full expression of the scattering matrix becomes quite lengthy, we
list its elements in Appendix~\ref{app:two-points}. Using
Eqs.~\eqref{eq:Stt-psi-psi} and~\eqref{eq:Stt-psid-psi}, we obtain the kernel
defined in Eq.~\eqref{eq:K} as
\begin{multline}
\label{eq:K-tt}
K(E)= 1- \cos (2 |\tilde{t}_1|) \cos (2 |\tilde{t}_2|)
+ \sin (2 |\tilde{t}_1|) \sin (2 |\tilde{t}_2|)
\\
\times \left(\cos^2( |\tilde{t}_1|) \cos(\tilde{k} a -\phi_{12} )
+\sin^2(|\tilde{t}_1|) \cos (\tilde{k} a + \phi_{12})\right).
\end{multline}
Here, we have used the notation $\tilde{k} =k_f -k_m$ and
$\phi_{12}=\phi_{1}-\phi_{2}$ and defined effective tunneling constants
$|\tilde{t}_i|={ |t_i|}/{\left(2 \sqrt{v_\text{f} v_\text{m}}\right)}$,
$i=1,2$. To have non-vanishing thermopower, $K(E)$ needs to be a non-even
function of energy, which requires the phase $\alpha_1\neq \alpha_2 +n \pi$.
Since the phase difference of the contact tunneling strengths can be
arbitrary (or tuned by threading magnetic flux), an electric voltage
difference will generically appear between the conducting lead and the
Majorana edge state. We then expect that a finite tunneling region, in
general, leads to a non-vanishing Seebeck coefficient.

We conclude this section by discussing possible signatures of the Seebeck
effect in the double-point-contact setup. The purpose is to show that a
non-vanishing Seebeck coefficient of reasonable magnitude can appear. In the
weak tunneling limit, $ |\tilde{t}_1|\sim |\tilde{t}_2| \ll 1$, the energy
derivative of the differential conductance can be evaluated from
Eq.~\eqref{eq:K-tt}, and is given by
\begin{equation}
\frac{d \ln G(E)}{dE}\big|_{E=0} \approx \frac{a}{\hbar}
\frac{\sin\phi_{12}}{1+\cos\phi_{12} }
\left(\frac{1}{v_\text{f}} -\frac{1}{v_\text{m}} \right).
\end{equation}
From the Mott formula in Eq.~\eqref{eq:Seebeck-coeff}, the Seebeck
coefficient can be approximated to be
\begin{equation}
\mathcal{S} \approx -  \frac{\pi^2}{3} \frac{k_{\text B}^2 T_s}{e}
\frac{a}{\hbar  v_\text{m}}.
\end{equation}
where we have used $v_\text{m}\ll v_\text{f}$ as the velocity of the Majorana edge state is reduced by comparison with the Fermi velocity,~\cite{Fu09} and we have taken the phase-dependent factor, $\sin(\phi_{12})/(1+\cos(\phi_{12}))\sim
1$, as its median value. For $a \sim 0.1-10\; \mu$m and $v_\text{m}\sim 10^4$
m/s at $T_s\sim 100$~mK, we expect the value of the thermopower
\begin{equation}
\label{eq:S-numeric}
\mathcal{S} \sim 10^{-5} - 10^{-3} \; \text{V/K}.
\end{equation}
This will result in a reasonably strong signal for the potential difference
$\Delta V$ when the temperature difference $\Delta T$ is around $1\sim 10$
mK. In this limit, we note that the Seebeck coefficient scales linearly with
the distance between the two contacts while it is inversely proportional to
the propagating velocity of the Majorana edge mode.

\section{Conclusions}
\label{sec:conclusion}

In summary, by employing the Landauer-B\" uttiker formalism, we have
demonstrated explicitly that the thermopower (Seebeck coefficient) between a
conducting lead and a superconductor satisfies the Mott formula. Using point-contact models, we argued that the thermopower between a conducting lead and
a Majorana edge state generically does not vanish. In the absence of current,
this leads to a finite voltage when a temperature difference is established
across a tunneling junction. With the aid of the Mott formula, the
temperature of the Majorana edge state can be inferred by measuring the
differential conductance and the voltage across the tunneling region in the
absence of current flow. Since this technique has been demonstrated in
non-superconducting systems for tunneling
geometries~\cite{Molenkamp90,Appleyard98,Granger09}, we expect that a similar
technique can be used to probe the temperature of Majorana edge states.

\section*{Acknowledgments}
The authors would like to thank A.~R.~Akhmerov and D.~Pekker for helpful
discussions. CYH and KS were supported in part by the DARPA-QuEST program. KS
was supported in part by NSF award DMR-0748925. GR is grateful for support
from the Packard foundation and the IQIM, an NSF center supported in part by the Moore fundation.

\appendix

\section{Proof of the Onsager relation}
\label{app:Onsager-relation}

\subsection{Time-reversal-invariant case}

From Eq.~\eqref{eq:def-Lij}, the form of the following linear-response coefficients
\begin{align}
\label{eq:L12-app}
L_{12} =& \frac{1}{h} \int_{-\infty}^{\infty} d E\, E
\left(- \frac{\partial f}{\partial E} \right)
\left( N_{np} - P_{np}^{np}+ P_{np}^{nh} \right),
\\
\label{eq:L21-app}
L_{21} =& \frac{1}{h} \int_{-\infty}^{\infty} d E\, E
\left(- \frac{\partial f}{\partial E} \right)
\left( N_{np} - P_{np}^{np}- P_{np}^{nh} \right),
\end{align}
is not identical and does not obviously satisfy the Onsager relation. Our
goal here is to show that these linear response coefficients do in fact
follow the Onsager reciprocity relation $L_{12}=L_{21}$ in the presence of
time reversal symmetry (TRS). We observe that $L_{12}$ and $L_{21}$  differ
only by a minus sign in the last term of the integrand. Hence, to satisfy the
Onsager relation, the contribution from the last term has to vanish after the
integration. In the presence of TRS, we will show that $P_{np}^{nh}(E)$ is
indeed an even function of energy and hence does not contribute to the
integrals in Eqs.~(\ref{eq:L12-app},\ref{eq:L21-app}).

Let us define the basis for a scattering problem between a normal and a
superconducting lead for spin-$1/2$ electrons as
\begin{multline}
\label{eq:basis-app}
\left( \psi_{np\uparrow}(E), \psi_{sp\uparrow}(E), \psi_{np\downarrow}(E),
\psi_{sp\downarrow}(E),\right.
\\
\left. \psi_{nh\uparrow}(E), \psi_{sh\uparrow}(E), \psi_{nh\downarrow}(E),
\psi_{sh\downarrow}(E) \right)^{T},
\end{multline}
where each $\psi_{i \alpha \sigma}(E)$ has $N_{i \alpha\sigma}(E)$ channels
that will be indicated by indices $a/b$ in what follows, and $\sigma$ is the
spin index. In this basis, a scattering matrix $S(E)$ has matrix elements
\begin{equation}
\label{eq:S-E}
S(E)=\left\{ s_{j \beta \sigma'}^{i \alpha \sigma}(E)\right\}_{i,j=n/s; \alpha,\beta= p/h; \sigma,\sigma'=\uparrow,\downarrow}
\end{equation}
where each $s_{j \beta \sigma'}^{\ell \alpha \sigma}$ is an $N_{\ell\alpha
\sigma}(E)\times N_{j\beta \sigma'}(E)$ matrix with its elements denoted by
$s^{\ell\alpha \sigma a}_{j \beta \sigma' b}(E)$, which relates the outgoing
current at state $(\ell,\alpha, \sigma',b)$ to the incoming current at state
$(j,\beta, \sigma, a)$.

For spin-1/2 electrons, the time reversal transformation of the scattering
matrix in the electron basis is given by
\begin{equation}
\label{eq:def-TRS-app}
\mathcal{T} [S] = \sigma_y S^{T} \sigma_y
\end{equation}
where the $\sigma_y$ is the Pauli matrix acting on spins and the superscript $T$ indicates the transpose of the matrix. With the TRS,
we have $\mathcal{T}[S]=S$. (A similar transformation can be defined for spinless electrons by $\mathcal{T} [S] = S^{T} $. Then all conclusions in
this appendix will follow.) For a superconductor, the corresponding time reversal transformation becomes
$\mathcal{T}[S]= (\openone_{ph} \otimes \sigma_y \otimes \openone) S^T
(\openone_{ph} \otimes \sigma_y \otimes \openone )$, where the
$\openone_{ph}$ acts on the particle--hole indices, $\sigma_y$ acts on the
spin indices, and $\openone$ acts on the lead$\times$channel indices. The
time reversal transformation maps matrix elements of the scattering matrix by
\begin{equation}
\label{eq:T-S-E}
\mathcal{T}[S(E)] = \left\{ s_{j \beta \sigma'}^{\ell \alpha \sigma}(E)
\mapsto (\sigma \sigma') \times s^{j \beta \bar{\sigma}'}_{i \alpha \bar{\sigma} }
(E)^{T} \right\}
\end{equation}
where $\bar{\sigma}$ and $\bar{\sigma}'$ indicate the flip of spin, i.e.,
$\bar{\uparrow}= \downarrow$ and vise versa, and the values of
$\sigma,\sigma'$ inside the parentheses are taken to be $\pm 1$ for
$\sigma,\sigma'=\uparrow/\downarrow$.

The presence of TRS implies $S(E)= \mathcal{T}[S(E)]$ and leads to following
useful identities for each element
\begin{equation}
\label{eq:TRS-identities-1}
\begin{split}
s^{n h \uparrow a }_{n p \uparrow b} (E) = s^{ n p \downarrow b}_{n h \downarrow a} (E), & \quad
 s^{n h \downarrow a}_{ np \downarrow b}(E) = s^{n p \uparrow b}_{n h \uparrow a} (E) ,
\\
s^{n h \uparrow a}_{n p \downarrow b} (E) = - s^{n p \uparrow b}_{n h\downarrow a} (E), & \quad
s^{n h\downarrow a}_{n p\uparrow b}  (E)= - s^{n p\downarrow b}_{n h\uparrow a} (E) .
\end{split}
\end{equation}
Let us recall the definition of the scattering probability $P_{np}^{nh}(E)$
in terms of of the scattering matrix elements
\begin{equation}
\label{eq:P11-+-app}
\begin{split}
P_{np}^{nh} (E) = & \sum_{\sigma,\sigma';a,b} s^{n h \sigma a}_{n p \sigma' b} (E) s^{n h\sigma a}_{n p \sigma' b} (E)^*
\\
=& - \sum_{\sigma,\sigma';a,b} s^{n h \sigma a}_{n p\sigma' b} (E) s^{n p \sigma a}_{n h \sigma' b} (-E) .
\end{split}
\end{equation}
where we have used the particle--hole symmetry (PHS) given by
Eq.~\eqref{eq:s-matrix-relation-PH-sym} for the second equality. With the aid
of the identities given by Eq.~\eqref{eq:TRS-identities-1}, we have
\begin{equation}
\label{eq:T-reversed-P11-+}
\begin{split}
P_{np}^{nh} (E) = & \sum_{\sigma,\sigma';a,b} s^{n p \sigma b}_{n h \sigma' a}(E)
s^{n p \sigma b}_{n h \sigma' a} (E)^*
\\
=& - \sum_{\sigma,\sigma';a,b} s^{n p \sigma a}_{n h \sigma'  b} (E)
s^{n h \sigma a}_{n p \sigma' b} (-E),
\end{split}
\end{equation}
where we again used the PHS for the last equality. Comparing the results in Eqs.~\eqref{eq:P11-+-app} and \eqref{eq:T-reversed-P11-+}, we can conclude
that $P^{nh}_{np} (E)= P^{nh}_{np} (-E)$. Hence, $P_{np}^{nh} (E)$ is an even
function of the energy $E$ and leads to no contribution of the linear
response coefficient. Thus, the Onsager reciprocal relation is satisfied in
the presence of TRS.

\subsection{Generic case}

In the absence of TRS, the Onsager reciprocal relation states that $L_{ij} =
\mathcal{T}[L_{ji}]$, where $\mathcal{T}[L_{ji}]$ stands for the linear-response coefficient of the time-reversed system. Therefore we need to verify
that the following relations are satisfied
\begin{equation}
L_{11}=\mathcal{T}[L_{11}] , \quad L_{22}=\mathcal{T}[L_{22}], \quad L_{12}=\mathcal{T}[L_{21}].
\end{equation}
As Eqs.~\eqref{eq:def-Lij} involve one channel number and two tunneling
probabilities, $N_{n p}(E)$, $P_{np}^{np}(E)$ and $P_{np}^{nh}(E)$, we shall
focus on those quantities of the time-reversed system.

First, the number of channels is invariant under the time reversal, i.e.,
$\mathcal{T}[N_{np}(E)] = N_{np}(E)$. Second, from the definition of the
scattering probability $ P^{np}_{np}(E) = \sum_{\sigma,\sigma';a, b} |s^{n p
\sigma a}_{n p \sigma' b}|^2, $ and time reversed elements in
Eq.~\eqref{eq:T-S-E}, $\mathcal{T}[s^{n p \sigma}_{n p \sigma'}] = (\sigma
\sigma')\times {s^{n p \bar{\sigma}'}_{n p \bar{\sigma}}}^T$, we have
$\mathcal{T}[P_{np}^{np}(E)] = P_{np}^{np}(E)$ under the time-reversed transformation. Finally, the time reversed form of
$\mathcal{T}[P_{np}^{nh}(E)] =  \sum_{\sigma,\sigma';a,b} |s^{n p \sigma
a}_{n h \sigma' b} (E)|^2$, given in Eq.~\eqref{eq:T-reversed-P11-+}, is not
invariant under the time reversal transformation.

We shall now discuss each Onsager relation separately.

\subsubsection{Diagonal response coefficients}

Let us recall the linear response coefficient in Eq.~\eqref{eq:def-Lij}
\begin{equation}
L_{11} = \frac{1}{h} \int_{-\infty}^{\infty} d E \left(- \frac{\partial f}{\partial E} \right)  \left( N_{n p} - P_{np}^{np}+ P_{np}^{nh} \right).
\end{equation}
As both $N_{np}(E)$ and $P_{np}^{np}(E)$ are invariant under the time reversal transformation, we have

\begin{equation}
\begin{split}
&L_{11} -\mathcal{T}[L_{11} ]
\\
=&  \frac{1}{h} \int d E \left(- \frac{\partial f}{\partial E} \right) \left( P_{np}^{nh}  - \mathcal{T}[P_{np}^{nh}] \right)
\\
=& \frac{1}{h}  \int d E \left(- \frac{\partial f }{\partial E} \right) \sum_{\sigma,\sigma';a,b} \left( |s^{n h \sigma a}_{n p\sigma' b} |^2 - |s^{np\sigma b}_{nh\sigma' a} |^2 \right),
\\
=& \frac{-1}{h} \int_{-\infty}^{\infty} dE \left(- \frac{\partial f}{\partial E} \right) \sum_{\sigma,\sigma';a,b}\left( s^{n h \sigma a}_{n p\sigma' b} (E) s^{n p \sigma a}_{nh \sigma' b} (-E) \right.
\\
&\qquad \qquad \qquad \qquad \qquad -\left. s^{n p \sigma b}_{n h \sigma' a} (E) s^{n h \sigma b}_{n p \sigma' a} (- E) \right),
\end{split}
\end{equation}
where we have use the PHS for the last equality. As all terms inside
parentheses are, in overall, odd functions and $\left(- \frac{\partial
f(E)}{\partial E} \right)$ is an even function of energy, the integration
vanishes. Hence we have shown that $L_{11}=\mathcal{T}[L_{11}]$. A similar
argument shows that $L_{22}=\mathcal{T}[L_{22}]$ as well.

\subsubsection{Off-diagonal response coefficients}

From Eqs.~\eqref{eq:L12-app} and~\eqref{eq:L21-app}, we immediately have
\begin{equation}
\begin{split}
&L_{12} - \mathcal{T}[L_{21} ]
\\
=& \frac{1}{h} \int d E\,  E \left(- \frac{\partial f}{\partial E} \right) \left( P_{np}^{nh}  + \mathcal{T}[P_{np}^{nh}] \right)
\\
=&\frac{1}{h}  \int d E\,  E \left(- \frac{\partial f}{\partial E} \right) \sum_{\sigma,\sigma';a,b} \left( |s^{n h \sigma a}_{n p\sigma' b}|^2 +|s^{n p \sigma b}_{n h \sigma' a} |^2 \right),
\\
=&- \int_{-\infty}^{\infty}  d E\,  E \left(-\frac{\partial f}{\partial E} \right) \sum_{\sigma,\sigma';a,b}\left( s^{n h \sigma a}_{n p \sigma' b} (E) s^{n p \sigma a}_{n h \sigma' b} (-E) \right.
\\
&\qquad \qquad \qquad \qquad \qquad \quad +\left. s^{n p \sigma b}_{n h\sigma' a} (E) s^{n h\sigma b}_{np\sigma' a} (- E) \right)
\end{split}
\end{equation}
where we have used the PHS for the last equality. Since all terms inside the
parentheses are even functions of $E$ while $E \left(- {\partial
f(E)}/{\partial E} \right)$ is an odd function of energy, the integral
vanishes. We therefore obtain $L^{A}_{12}=\mathcal{T}[L^{A}_{21}]$.

\section{Scattering matrix elements of two point contact setup}
\label{app:two-points}

In this Appendix, we list scattering matrix elements $s_{\beta}^{\alpha}$ of
the two-points contact setup. The tunneling amplitudes are defined by
$t_1=|t_1|e^{i \phi_1}$ and $t_{2}= |t_2| e^{i \phi_2}$. It is convenient to
define effective tunneling amplitudes $|\tilde{t}_i|={ |t_i|}/{\left(2
\sqrt{v_\text{f} v_\text{m}}\right)}$. First, matrix elements associating
with the kernel are given by
\begin{widetext}
\begin{align}
\label{eq:Stt-psi-psi}
s^{p}_{p}=& \frac{1}{2}  \left(2 e^{i k_f a} \left( e^{2 i (\phi_1- \phi_2)} \sin ^2(|\tilde{t}_1|) \sin ^2(|\tilde{t}_2|)+  \cos ^2(|\tilde{t}_1|) \cos ^2(|\tilde{t}_2|) \right)- e^{i (k_m a+ \phi_1- \phi_2)} \sin (2 |\tilde{t}_1|) \sin (2|\tilde{t}_2|) \right),
\\
\label{eq:Stt-psid-psi}
s^{h}_{p} = & - \frac{1}{2} \left( 2 e^{i k_f a} \left(e^{2 i \phi_1} \sin^2( |\tilde{t}_1|) \cos^2 (|\tilde{t}_2|) + e^{2 i \phi_2} \cos^2 (|\tilde{t}_1|) \sin^2(|\tilde{t}_2|)\right) + e^{i (k_m a+ \phi_1+\phi_2)}  \sin (2 |\tilde{t}_1|) \sin (2 |\tilde{t}_2|) \right).
\end{align}
The rest of the matrix elements are given by
\begin{align}
\label{eq:Stt-eta-eta}
s^{\eta}_{\eta}=&e^{i k_m a} \cos (2|\tilde{t}_1|) \cos(2|\tilde{t}_2|)-e^{i k_f a} \cos ( \phi_1-  \phi_2) \sin(2|\tilde{t}_1|) \sin(2|\tilde{t}_2|)
 \\
s^{\eta}_{p}=& - \frac{1}{\sqrt{2}} \left( e^{i (k_f a- \phi_2 )} \sin(2|\tilde{t}_2|) \left( e^{2 i  \phi_2 } \cos^2( |\tilde{t}_1|)-e^{2 i  \phi_1} \sin^2( |\tilde{t}_1| \right)) + e^{i (k_m a+ \phi_1)} \sin(2 |\tilde{t}_1|) \cos(2 |\tilde{t}_2|)\right)
\\
s^{\eta}_{h}=& - \frac{1}{\sqrt{2}} \left(e^{i (k_f a+  \phi_2)} \sin (2 |\tilde{t}_2|) \left(e^{-2 i  \phi_2} \cos^2(|\tilde{t}_2|)-e^{-2 i  \phi_1} \sin ^2(|\tilde{t}_1|)\right)+e^{i (k_m a- \phi_1)} \sin (2 |\tilde{t}_1|) \cos (2 |\tilde{t}_2|) \right)
\\
s^{p}_{\eta}=&  \frac{1}{\sqrt{2}} \left(e^{i (k_f a +  \phi_1) } \sin (2 |\tilde{t}_1|) \left( e^{ - 2 i  \phi_1 } \cos^2(|\tilde{t}_2|)-e^{ - 2 i  \phi_2} \sin ^2(|\tilde{t}_2|)\right)+ e^{i ( k_m a-  \phi_2)} \cos (2 |\tilde{t}_1|) \sin (2 |\tilde{t}_2|) \right)
\\
s^{p}_{h}=& - \frac{1}{2} \left( 2 e^{i k_f a} \left(e^{-2 i  \phi_1} \sin^2( |\tilde{t}_1|) \cos^2 (|\tilde{t}_2|) + e^{-2 i \phi_2} \cos^2 (|\tilde{t}_1|) \sin^2(|\tilde{t}_2|)\right) + e^{i (k_m a-  \phi_1- \phi_2)}  \sin (2 |\tilde{t}_1|) \sin (2 |\tilde{t}_2|) \right)
\\
s^{h}_{\eta} = & \frac{1}{\sqrt{2}} \left( e^{i (k_f a- \phi_1) } \sin (2 |\tilde{t}_1|) \left(e^{2 i  \phi_1} \cos^2 (|\tilde{t}_2|)-e^{ 2 i  \phi_2 } \sin ^2(|\tilde{t}_2|) \right)+e^{i (k_m a +  \phi_2)} \cos(2 |\tilde{t}_1|) \sin (2|\tilde{t}_2|) \right)
\\
\label{eq:Stt-psid-psid}
s^{h}_{h} = & \frac{1}{2} \left(2 e^{i k_f a} \left( e^{-2 i ( \phi_1- \phi_2)}  \sin^2 ( |\tilde{t}_1|) \sin^2 ( |\tilde{t}_2|) + \cos^2( |\tilde{t}_1|) \cos ^2( |\tilde{t}_2|) \right) - e^{i ( k_m a- \phi_1+  \phi_2)} \sin (2  |\tilde{t}_1|) \sin (2  |\tilde{t}_2|) \right)
\end{align}
\end{widetext}
With these matrix elements, one can show that the scattering matrix
$S^{tt}(\mathcal{E})$ is unitary and satisfies the particle--hole symmetry
defined in Eq.~\eqref{eq:s-matrix-relation-PH-sym}.

\bibliographystyle{apsrev}
\bibliography{thermoelectric}

\begin{thebibliography}{24}
\expandafter\ifx\csname natexlab\endcsname\relax\def\natexlab#1{#1}\fi
\expandafter\ifx\csname bibnamefont\endcsname\relax
  \def\bibnamefont#1{#1}\fi
\expandafter\ifx\csname bibfnamefont\endcsname\relax
  \def\bibfnamefont#1{#1}\fi
\expandafter\ifx\csname citenamefont\endcsname\relax
  \def\citenamefont#1{#1}\fi
\expandafter\ifx\csname url\endcsname\relax
  \def\url#1{\texttt{#1}}\fi
\expandafter\ifx\csname urlprefix\endcsname\relax\def\urlprefix{URL }\fi
\providecommand{\bibinfo}[2]{#2}
\providecommand{\eprint}[2][]{\url{#2}}

\bibitem[{\citenamefont{Kane and Fisher}(1996)}]{Kane96}
\bibinfo{author}{\bibfnamefont{C.~L.} \bibnamefont{Kane}} \bibnamefont{and}
  \bibinfo{author}{\bibfnamefont{M.~P.~A.} \bibnamefont{Fisher}},
  \bibinfo{journal}{Phys. Rev. Lett.} \textbf{\bibinfo{volume}{76}},
  \bibinfo{pages}{3192} (\bibinfo{year}{1996}).

\bibitem[{\citenamefont{Kane and Fisher}(1997)}]{Kane97}
\bibinfo{author}{\bibfnamefont{C.~L.} \bibnamefont{Kane}} \bibnamefont{and}
  \bibinfo{author}{\bibfnamefont{M.~P.~A.} \bibnamefont{Fisher}},
  \bibinfo{journal}{Phys. Rev. B} \textbf{\bibinfo{volume}{55}},
  \bibinfo{pages}{15832} (\bibinfo{year}{1997}).

\bibitem[{\citenamefont{Mott and Jones}(1958)}]{Mott}
\bibinfo{author}{\bibfnamefont{N.~F.} \bibnamefont{Mott}} \bibnamefont{and}
  \bibinfo{author}{\bibfnamefont{H.}~\bibnamefont{Jones}},
  \emph{\bibinfo{title}{The Theory of the Properties of Metals and Alloys}}
  (\bibinfo{publisher}{Dover Publications, New York}, \bibinfo{year}{1958}).

\bibitem[{\citenamefont{Jonson and Mahan}(1980)}]{Jonson80}
\bibinfo{author}{\bibfnamefont{M.}~\bibnamefont{Jonson}} \bibnamefont{and}
  \bibinfo{author}{\bibfnamefont{G.~D.} \bibnamefont{Mahan}},
  \bibinfo{journal}{Phys. Rev. B} \textbf{\bibinfo{volume}{21}},
  \bibinfo{pages}{4223} (\bibinfo{year}{1980}).

\bibitem[{\citenamefont{Sivan and Imry}(1986)}]{Sivan86}
\bibinfo{author}{\bibfnamefont{U.}~\bibnamefont{Sivan}} \bibnamefont{and}
  \bibinfo{author}{\bibfnamefont{Y.}~\bibnamefont{Imry}},
  \bibinfo{journal}{Phys. Rev. B} \textbf{\bibinfo{volume}{33}},
  \bibinfo{pages}{551} (\bibinfo{year}{1986}).

\bibitem[{\citenamefont{Molenkamp et~al.}(1990)\citenamefont{Molenkamp, van
  Houten, Beenakker, Eppenga, and Foxon}}]{Molenkamp90}
\bibinfo{author}{\bibfnamefont{L.~W.} \bibnamefont{Molenkamp}},
  \bibinfo{author}{\bibfnamefont{H.}~\bibnamefont{van Houten}},
  \bibinfo{author}{\bibfnamefont{C.~W.~J.} \bibnamefont{Beenakker}},
  \bibinfo{author}{\bibfnamefont{R.}~\bibnamefont{Eppenga}}, \bibnamefont{and}
  \bibinfo{author}{\bibfnamefont{C.~T.} \bibnamefont{Foxon}},
  \bibinfo{journal}{Phys. Rev. Lett.} \textbf{\bibinfo{volume}{65}},
  \bibinfo{pages}{1052} (\bibinfo{year}{1990}).

\bibitem[{\citenamefont{Appleyard et~al.}(1998)\citenamefont{Appleyard,
  Nicholls, Simmons, Tribe, and Pepper}}]{Appleyard98}
\bibinfo{author}{\bibfnamefont{N.~J.} \bibnamefont{Appleyard}},
  \bibinfo{author}{\bibfnamefont{J.~T.} \bibnamefont{Nicholls}},
  \bibinfo{author}{\bibfnamefont{M.~Y.} \bibnamefont{Simmons}},
  \bibinfo{author}{\bibfnamefont{W.~R.} \bibnamefont{Tribe}}, \bibnamefont{and}
  \bibinfo{author}{\bibfnamefont{M.}~\bibnamefont{Pepper}},
  \bibinfo{journal}{Phys. Rev. Lett.} \textbf{\bibinfo{volume}{81}},
  \bibinfo{pages}{3491} (\bibinfo{year}{1998}).

\bibitem[{\citenamefont{Granger et~al.}(2009)\citenamefont{Granger, Eisenstein,
  and Reno}}]{Granger09}
\bibinfo{author}{\bibfnamefont{G.}~\bibnamefont{Granger}},
  \bibinfo{author}{\bibfnamefont{J.~P.} \bibnamefont{Eisenstein}},
  \bibnamefont{and} \bibinfo{author}{\bibfnamefont{J.~L.} \bibnamefont{Reno}},
  \bibinfo{journal}{Phys. Rev. Lett.} \textbf{\bibinfo{volume}{102}},
  \bibinfo{pages}{086803} (\bibinfo{year}{2009}).

\bibitem[{\citenamefont{Read and Green}(2000)}]{Read00}
\bibinfo{author}{\bibfnamefont{N.}~\bibnamefont{Read}} \bibnamefont{and}
  \bibinfo{author}{\bibfnamefont{D.}~\bibnamefont{Green}},
  \bibinfo{journal}{Phys. Rev. B} \textbf{\bibinfo{volume}{61}},
  \bibinfo{pages}{10267} (\bibinfo{year}{2000}).

\bibitem[{\citenamefont{B\"uttiker}(1986)}]{Buttiker86}
\bibinfo{author}{\bibfnamefont{M.}~\bibnamefont{B\"uttiker}},
  \bibinfo{journal}{Phys. Rev. Lett.} \textbf{\bibinfo{volume}{57}},
  \bibinfo{pages}{1761} (\bibinfo{year}{1986}).

\bibitem[{\citenamefont{Datta}(1995)}]{Datta}
\bibinfo{author}{\bibfnamefont{S.}~\bibnamefont{Datta}},
  \emph{\bibinfo{title}{Electronic Transport in Mesoscopic Systems}}
  (\bibinfo{publisher}{Cambridge University Press, Cambridge},
  \bibinfo{year}{1995}).

\bibitem[{\citenamefont{Blonder et~al.}(1982)\citenamefont{Blonder, Tinkham,
  and Klapwijk}}]{Blonder82}
\bibinfo{author}{\bibfnamefont{G.~E.} \bibnamefont{Blonder}},
  \bibinfo{author}{\bibfnamefont{M.}~\bibnamefont{Tinkham}}, \bibnamefont{and}
  \bibinfo{author}{\bibfnamefont{T.~M.} \bibnamefont{Klapwijk}},
  \bibinfo{journal}{Phys. Rev. B} \textbf{\bibinfo{volume}{25}},
  \bibinfo{pages}{4515} (\bibinfo{year}{1982}).

\bibitem[{\citenamefont{Claughton and Lambert}(1996)}]{Claughton96}
\bibinfo{author}{\bibfnamefont{N.~R.} \bibnamefont{Claughton}}
  \bibnamefont{and} \bibinfo{author}{\bibfnamefont{C.~J.}
  \bibnamefont{Lambert}}, \bibinfo{journal}{Phys. Rev. B}
  \textbf{\bibinfo{volume}{53}}, \bibinfo{pages}{6605} (\bibinfo{year}{1996}).

\bibitem[{\citenamefont{Anantram and Datta}(1996)}]{Anantram96}
\bibinfo{author}{\bibfnamefont{M.~P.} \bibnamefont{Anantram}} \bibnamefont{and}
  \bibinfo{author}{\bibfnamefont{S.}~\bibnamefont{Datta}},
  \bibinfo{journal}{Phys. Rev. B} \textbf{\bibinfo{volume}{53}},
  \bibinfo{pages}{16390} (\bibinfo{year}{1996}).

\bibitem[{not({\natexlab{a}})}]{note1}
\bibinfo{note}{We should point out that a voltage across the interface between
  a normal metal and a superconductor in response to the temperature difference
  between the two regions had been observed in the
  past~\cite{Harlingen81,Waldram92}. However, it had not been established
  whether the Mott formula is applicable to such a setup.}

\bibitem[{\citenamefont{Hou et~al.}(2012)\citenamefont{Hou, Shtengel, Refael,
  and Goldbart}}]{Hou12}
\bibinfo{author}{\bibfnamefont{C.-Y.} \bibnamefont{Hou}},
  \bibinfo{author}{\bibfnamefont{K.}~\bibnamefont{Shtengel}},
  \bibinfo{author}{\bibfnamefont{G.}~\bibnamefont{Refael}}, \bibnamefont{and}
  \bibinfo{author}{\bibfnamefont{P.~M.} \bibnamefont{Goldbart}},
  \bibinfo{journal}{New Journal of Physics} \textbf{\bibinfo{volume}{14}},
  \bibinfo{pages}{105005} (\bibinfo{year}{2012}).

\bibitem[{\citenamefont{Reimann and Manninen}(2002)}]{Reimann02}
\bibinfo{author}{\bibfnamefont{S.~M.} \bibnamefont{Reimann}} \bibnamefont{and}
  \bibinfo{author}{\bibfnamefont{M.}~\bibnamefont{Manninen}},
  \bibinfo{journal}{Rev. Mod. Phys.} \textbf{\bibinfo{volume}{74}},
  \bibinfo{pages}{1283} (\bibinfo{year}{2002}).

\bibitem[{\citenamefont{Law et~al.}(2009)\citenamefont{Law, Lee, and
  Ng}}]{Law09}
\bibinfo{author}{\bibfnamefont{K.~T.} \bibnamefont{Law}},
  \bibinfo{author}{\bibfnamefont{P.~A.} \bibnamefont{Lee}}, \bibnamefont{and}
  \bibinfo{author}{\bibfnamefont{T.~K.} \bibnamefont{Ng}},
  \bibinfo{journal}{Phys. Rev. Lett.} \textbf{\bibinfo{volume}{103}},
  \bibinfo{pages}{237001} (\bibinfo{year}{2009}).

\bibitem[{not({\natexlab{b}})}]{note2}
\bibinfo{note}{Chiral Majorana edge modes considered here should not be
  confused with Majorana \emph{zero} modes; in our case these modes are formed
  out of the gapless states within the superconducting gap. Strictly speaking,
  these Majorana modes are not even energy eigenstates as they mix $\pm E$
  states.}

\bibitem[{\citenamefont{Ashcroft and Mermin}(1976)}]{Ashcroft}
\bibinfo{author}{\bibfnamefont{N.~W.} \bibnamefont{Ashcroft}} \bibnamefont{and}
  \bibinfo{author}{\bibfnamefont{N.~D.} \bibnamefont{Mermin}},
  \emph{\bibinfo{title}{Solid State Physics}} (\bibinfo{publisher}{Harcourt
  College Publishers, New York}, \bibinfo{year}{1976}).

\bibitem[{\citenamefont{Li et~al.}(2012)\citenamefont{Li, Fleury, and
  B\"uttiker}}]{Li12}
\bibinfo{author}{\bibfnamefont{J.}~\bibnamefont{Li}},
  \bibinfo{author}{\bibfnamefont{G.}~\bibnamefont{Fleury}}, \bibnamefont{and}
  \bibinfo{author}{\bibfnamefont{M.}~\bibnamefont{B\"uttiker}},
  \bibinfo{journal}{Phys. Rev. B} \textbf{\bibinfo{volume}{85}},
  \bibinfo{pages}{125440} (\bibinfo{year}{2012}).

\bibitem[{\citenamefont{Fu and Kane}(2009)}]{Fu09}
\bibinfo{author}{\bibfnamefont{L.}~\bibnamefont{Fu}} \bibnamefont{and}
  \bibinfo{author}{\bibfnamefont{C.~L.} \bibnamefont{Kane}},
  \bibinfo{journal}{Phys. Rev. Lett.} \textbf{\bibinfo{volume}{102}},
  \bibinfo{pages}{216403} (\bibinfo{year}{2009}).

\bibitem[{\citenamefont{Harlingen}(1981)}]{Harlingen81}
\bibinfo{author}{\bibfnamefont{D.}~\bibnamefont{Harlingen}},
  \bibinfo{journal}{Journal of Low Temperature Physics}
  \textbf{\bibinfo{volume}{44}}, \bibinfo{pages}{163} (\bibinfo{year}{1981}).

\bibitem[{\citenamefont{Waldram and Battersby}(1992)}]{Waldram92}
\bibinfo{author}{\bibfnamefont{J.~R.} \bibnamefont{Waldram}} \bibnamefont{and}
  \bibinfo{author}{\bibfnamefont{S.~J.} \bibnamefont{Battersby}},
  \bibinfo{journal}{J. Low Temp. Phys.} \textbf{\bibinfo{volume}{86}},
  \bibinfo{pages}{1} (\bibinfo{year}{1992}).

\end{thebibliography}

\end{document}